\documentclass[twocolumn]{article}
\usepackage[top=2cm, bottom=2cm, left=1.5cm, right=1.5cm]{geometry}

\usepackage[colorinlistoftodos]{todonotes}
\usepackage{authblk}
\usepackage{hyperref}
\usepackage{balance}
\usepackage{colortbl}
\usepackage{tabularx}

\title{A Preliminary Study for Identification of Additive Manufactured Objects with Transmitted Images}

\author[1]{Kenta Yamamoto\thanks{kenta.yam@digitalnature.slis.tsukuba.ac.jp}}
\author[1]{Ryota Kawamura}
\author[2]{Kazuki Takazawa}
\author[1]{Hiroyuki Osone}
\author[1,2]{Yoichi Ochiai}
\affil[1]{University of Tsukuba}
\affil[2]{Pixie Dust Technologies, Inc.}

\date{}

\begin{document}
\maketitle

\begin{abstract}
Additive manufacturing has the potential to become a standard method for manufacturing products, and product information is indispensable for the item distribution system.
While most products are given barcodes to the exterior surfaces, research on embedding barcodes inside products is underway.
This is because additive manufacturing makes it possible to carry out manufacturing and information adding at the same time, and embedding information inside does not impair the exterior appearance of the product.
However, products that have not been embedded information can not be identified, and embedded information can not be rewritten later.
In this study, we have developed a product identification system that does not require embedding barcodes inside.
This system uses a transmission image of the product which contains information of each product such as different inner support structures and manufacturing errors.
We have shown through experiments that if datasets of transmission images are available, objects can be identified with an accuracy of over 90\%.
This result suggests that our approach can be useful for identifying objects without embedded information.
\end{abstract}

\section{Introduction}
\label{sec:introduction}
The additive manufacturing (AM) process has greatly improved the ease of model customization. When managing products with a digital system, the addition of information is essential, such as a barcode. While management information has only to be attached to the exterior, additive manufacturing process makes it possible to embed information inside the product. Detection is then possible by various methods, for example, embedded information can be read using terahertz (Thz) sensing in InfraStructs\cite{Willis2013InfraStructsFI}, and using visible light projection in AirCode\cite{Li2017AirCodeUP}.

Two problems exist with the embedded information approach. The first is that an information adding step must be introduced into the manufacturing process. This also constitutes an advantage as the integration of manufacturing and adding information enables overall reduction of processing time and cost, nevertheless it is not possible to identify objects that were manufactured without embedded information. The second problem arises when the embedded information is duplicated such that product batches cannot be uniquely identified. This can occur in the current item distribution system. In previous research, defects in the barcode surface produced by the manufacturing process have been used for unique identification of products\cite{BarcodeUeno2018}.

In this study, a system of product identification was developed that does not use embedded information. All products have inherent structure combined with characteristics created by the manufacturing process, particularly manufacturing errors. Transmission images of product items are utilized to extract inherent product characteristics. This idea was inspired by research that uses the finger vein pattern to authenticate individuals. Finger vein pattern authentication is biometric, hence information is not added in advance of the authentication process; and, in order to visualize the internal blood vessels, an image of the finger transmitted by infrared light is used. Also in our system, information is not added, and we use images transmitted by infrared light.

To verify the identification system, two experiments were conducted: feature matching and deep learning; using two types of object group with identical exterior appearance. In one group, the internal structure of the object was varied. In the other group, all structures, including the internal structure, were unchanged; i.e., the objects differed only as a result of manufacturing errors.

The results suggest that feature matching is possible when there is a change in shape, and that deep learning can be used to identify objects that vary due to naturally occurring manufacturing errors. The purpose of this investigation is to examine the possibility of utilizing an identification system based on transmission images, to discriminate products in the absence of embedded information.

\begin{figure*}[t]
    \centering
    \includegraphics[width=\textwidth]{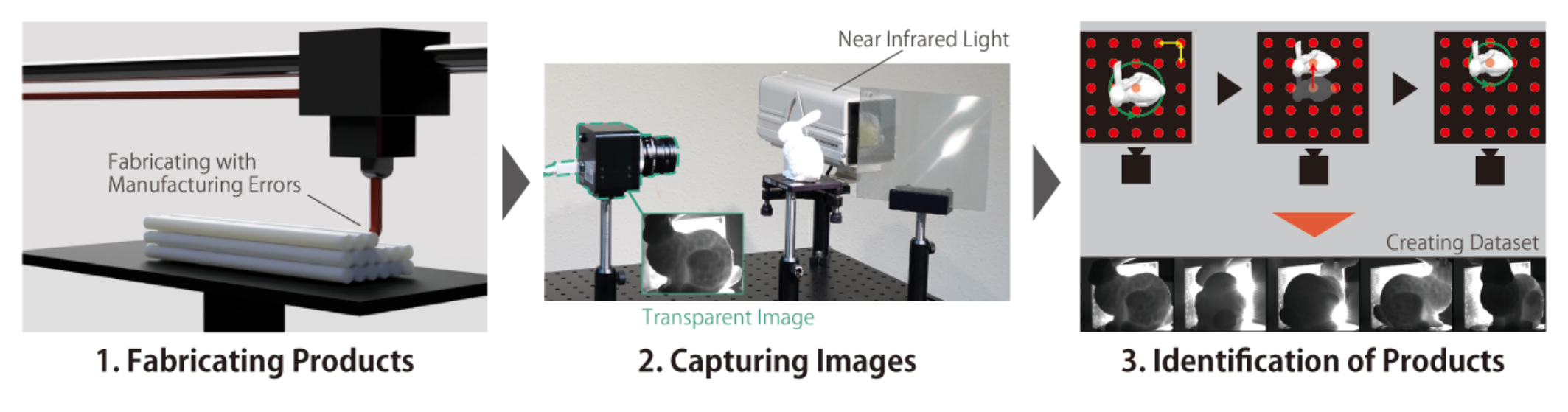}
    \caption{System Overview. The system is a three step process: fabrication, capturing, and identification. Products are fabricated without embedding information internally. Image capture requires transmitted light in the near infrared region. Identification is performed using feature matching or deep learning with datasets.}
    \label{fig:fig01}
\end{figure*}

\newcommand{\green}{\cellcolor[rgb]{0.65, 1.0, 0.65}}
\newcommand{\red}{\cellcolor[rgb]{1.0, 0.65, 0.65}}
\renewcommand{\arraystretch}{1.15}
\begin{table*}[t]
\footnotesize
\centering
\caption{Comparison with existing methods.}
\begin{tabular}{|l|c|c|c|c|c|c|}
\hline
& Information Place & Pre Embedding & Detection Angle & Learning Dataset & Wavelength & Reflection or Transmission \\ \hline
InfraStructs\cite{Willis2013InfraStructsFI}     & \red Inside            & \red Yes                    & \red One way         & \green No                      & THz        & Reflection                 \\ \hline
AirCode\cite{Li2017AirCodeUP} & \red Inside & \red Yes & \red One way & \green No & VIS & Reflection \\ \hline
LayerCode\cite{Maia2019LayerCodeOB} & \red Exterior & \red Yes & \green All angle & \green No & VIS or NIR & Reflection \\ \hline
Our System & \green Everywhere & \green No & \green All angle & \red Yes & NIR & Transmission \\ \hline
\end{tabular}
\end{table*}
\renewcommand{\arraystretch}{1.0}

\section{Related Works}
\label{sec:related_works}

\subsection{Properties of Additive Manufacturing}
Additive Manufacturing encompasses a variety of methods\cite{wong2012review}. Fused Deposition Modeling (FDM), in particular, has made a great contribution to the uptake of the desktop 3D printer. The widespread adoption of FDM has been accompanied by examination of the characteristics of FDM manufactured products. Many previous studies have considered how the characteristics of objects are affected by the manufacturing method itself; e.g., product characteristics\cite{FDMGrimm2003, Bakar2010AnalysisOF, Kumar2012ApplicationOF, Roberson20133DPS}, mechanical properties\cite{Lanzotti2015TheIO, Torres2016AnAF}, and material properties\cite{NovakovaMarcincinova2012BasicAA, MakerbotSchofield2015, MaterialSingh2017, ABSCantrell2017}. Several investigations have also been conducted to determine how objects change with different parameters; i.e., the relationship between each parameter and mechanical properties\cite{Torrado2016FailureAA, Rankouhi2016FailureAA, PLAChacon2017, FDMreviewPopescu2018}, and the correspondence between the internal structure and mechanical properties\cite{Baich2016StudyOI, FernandezVicenteMiguel2016EffectOI, KennyLAlvarez2016InvestigatingTI, Al2017ImprovingTS, AdditivelyRangisetty2017}. By referring to this body of literature concerning the FDM method, the fabrication characteristics of FDM could be established without the necessity to carry out exploratory research on the process.

\subsection{Biometrics using Transmitted Images}
It has become very common to perform personal authentication using biometric characteristics such as fingerprint authentication and face recognition. Most biometric authentication methods use an image produced by reflection of light from a target. However, some biometrics acquire an image after light has passed through the target. Finger vein pattern recognition is a typical example. In 2004, Miura et al.\cite{Miura2004} published a paper on personal authentication using a transmission image of a finger vein pattern obtained by illumination with infrared LEDs. The main challenge for this method lies in ensuring that the system is robust. It must be able to authenticate properly even if finger placement changes slightly, and it must not authenticate as another person. These problems led to research on how to automatically determine the region of interest\cite{miura2007extraction, Song2011AFV, Yang2012FingerveinRL} and improve imaging quality\cite{Tanaka2013DescatteringOT, Yang2014ComparativeCC}. Therefore, research literature concerning transparent biometric authentication was  referred to in order to assist with the construction of an identification system using transmission images of AM products.

\subsection{Embedding Information with Additive Manufacturing}
\label{sec:rw_embed}
Additive Manufacturing is a promising rapid prototyping method since it enables easy customization of shapes, thus prompting research on embedding barcodes during manufacturing. InfraStructs\cite{Willis2013InfraStructsFI} is a famous early study that embodies this concept. During the manufacturing process, a structure was built inside the object that was detected and decoded as information, using Thz sensing. This information could not be captured with a visible light camera. Several subsequent studies successfully embedded and read information in the object interior using alternative methods: i.e., internal information read in the far infrared region\cite{Okada2015NondestructivelyRO}, information embedded using highly reflective material\cite{Suzuki2017InformationHI}, and information embedded in the form of air pockets and read using visible light projection\cite{Li2017AirCodeUP}. Recently, Maia et al.\cite{Maia2019LayerCodeOB} embedded information in the entire surface of an object in order to simplify reading of the information.

Thus, existing research is based on building a system that embeds information in an invisible form during manufacturing and reads it out. In contrast, a method based on transmitted images does not require design of embedding information and the reading angle is free. Table 1 summarizes these differences. The utility of the transmission imaging system for identifying a product without preprocessing has been verified.
\renewcommand{\arraystretch}{1.2}
\begin{table}[t]
    \caption{Fabrication characteristics.}
    \footnotesize
    \label{tab:basic_feature}
    % \begin{tabular}{|c|l|l|} \hline
    % \begin{tabular*}{\linewidth}{@{\extracolsep{\fill}}|c|c|c|c|}\hline %ココ
    \begin{tabularx}{\linewidth}{|p{0.2\linewidth}|p{0.4\linewidth}|p{0.25\linewidth}|}\hline
        Main Topic &
        Sub Topic & 
        Sub-Sub Topic \\ \hline
         & Printing Width & \cellcolor[rgb]{0.9, 0.9, 0.9} \\ \cline{2-3}
         & Layer Thickness & \cellcolor[rgb]{0.9, 0.9, 0.9} \\ \cline{2-3}
        Shape & \red  & \red Infill Pattern \\ \cline{3-3}
        Parameters & \red Inner Support Structure & \red Infill Density \\ \cline{3-3}
         & \red & \red Infill Position \\ \hline

         &  & Material \\ \cline{3-3}
         & Material Properties & Color \\ \cline{2-3}
         & Printing Speed & \cellcolor[rgb]{0.9, 0.9, 0.9} \\ \cline{2-3}
        Condition & Temperature of Heat-Bed & \cellcolor[rgb]{0.9, 0.9, 0.9} \\ \cline{2-3}
        Parameters & Temperature of Printing Head & \cellcolor[rgb]{0.9, 0.9, 0.9} \\ \cline{2-3}
         & Retraction Speed & \cellcolor[rgb]{0.9, 0.9, 0.9} \\ \cline{2-3}
         
         & Environment & \cellcolor[rgb]{0.9, 0.9, 0.9} \\ \hline
    %  \end{tabular*}
    % \end{tabular}
    \end{tabularx}
\end{table}
\renewcommand{\arraystretch}{1.0}

\section{Materials and Methods}
\subsection{System Overview}
% \begin{figure*}[t]
%     \centering
%     \includegraphics[width=\textwidth]{fig/system_overview-v5.pdf}
%     \caption{System Overview. }
%     \label{fig:system_overview}
% \end{figure*}

This section details the system overview. The system consists of three processes: fabrication, capturing images, and identification (Fig.\ref{fig:fig01}). During fabrication, information is not intentionally embedded in products. Instead, inherent product characteristics are used for identification; i.e., visibly indistinguishable products may differ in their internal structure, and manufactured items hold intrinsic information such as dimensional values, which differ due to manufacturing errors. In the capturing process, transmission images are acquired. By using near infrared light, internal shape is captured in the transmitted image to enable identification of products on the basis of changes to internal structure. Inclusion of internal information also improves the accuracy of identifying changes resulting from manufacturing errors. Two methods were considered for the identification process: feature matching and deep learning. Feature matching does not require pre-learning for identification, allowing a simple system to be utilized. This contrasts with deep learning, which does require pre-learning, however discrimination accuracy is higher.

\subsection{Fabrication Characteristics}
Table 2 lists the characteristics that can appear in the transmission image of the product. All characteristics are referred to in the existing white paper\cite{FDMGrimm2003} and the software that runs the 3D printer. The characteristics were divided into two types, those related to shape and those related to condition. Of the shape-related properties, we paid particular attention to the inner support structure (see section 3.2.1). 
Many previous studies intentionally modified the interior space of a product without affecting the exterior appearance. In contrast, this study did not include intentional redesign of the interior space, but instead focused on random internal changes that occur naturally in manufacturing. Condition-related parameters were fixed in this study as it was difficult to distinguish between the effects of changing these parameters and differences arising from manufacturing errors. The assumption was made that if a manufacturing error could be identified, a change in condition parameters could also be distinguished. Therefore, only manufacturing errors were considered.

% [TODO] replace with High-Resolution PDF
\begin{figure}[t]
    \centering
    \includegraphics[width=\linewidth]{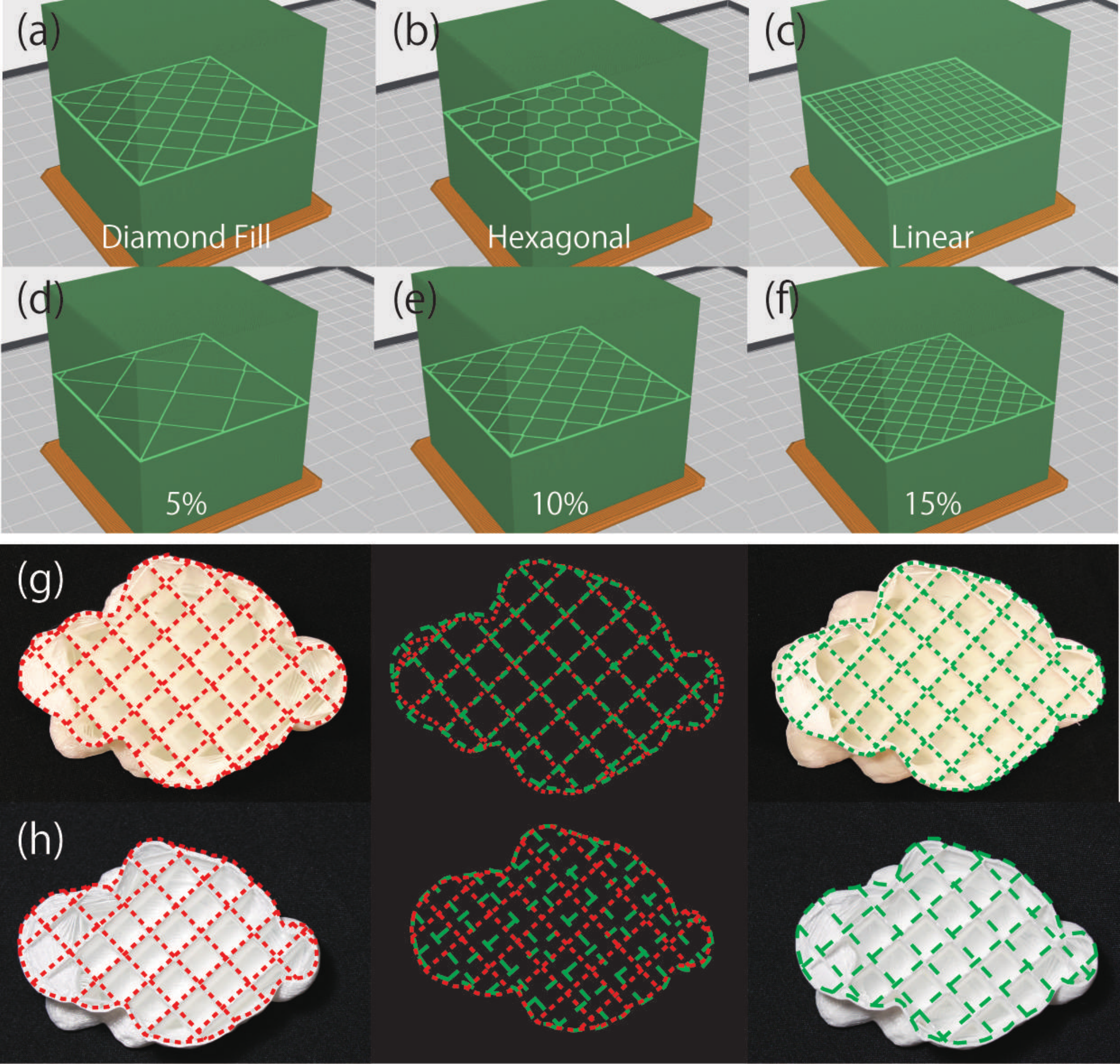}
    \caption{Shape parameter variations used in the identification tests. (a) to (c)  Different infill patterns. (d) to (f) Different infill densities. (g) and (h) Show the two in infill positions. (g) Initial infill position. (h) Alternative infill position.}
    \label{fig:fig02}
\end{figure}

{\bf Inner Support Structure:}
% \subsubsection{Inner Support Structure}
This parameter consists of three sub-parameters: infill pattern, infill density, and infill position (Fig.\ref{fig:fig02}), which are all visible on the transmission image. Infill pattern is an important parameter that is related to the shape of the internal structure and affects the mechanical properties of the product. Modifying the infill pattern changes the durability of the product under stress; i.e., the resistance to stress and the strength of the product are modified. Infill density refers to the density of a given infill pattern. Optimization of infill density is always desirable in terms of manufacturing cost and cycle time, as the amount of material increases with density. Infill position refers to the placement of the inner support structure for the given infill pattern. As placement is often done automatically by the slicer software, in general the system user is less aware of infill position.

\subsection{Transparent Imaging System}
The transmissive imaging system operates by capturing the characteristics and internal structure of products, both primitive and complex, in images using light wavelengths that are  transmitted through the target object. 

\subsubsection{Optical Setup}
\begin{figure}[t]
    \centering
    \includegraphics[width=\linewidth]{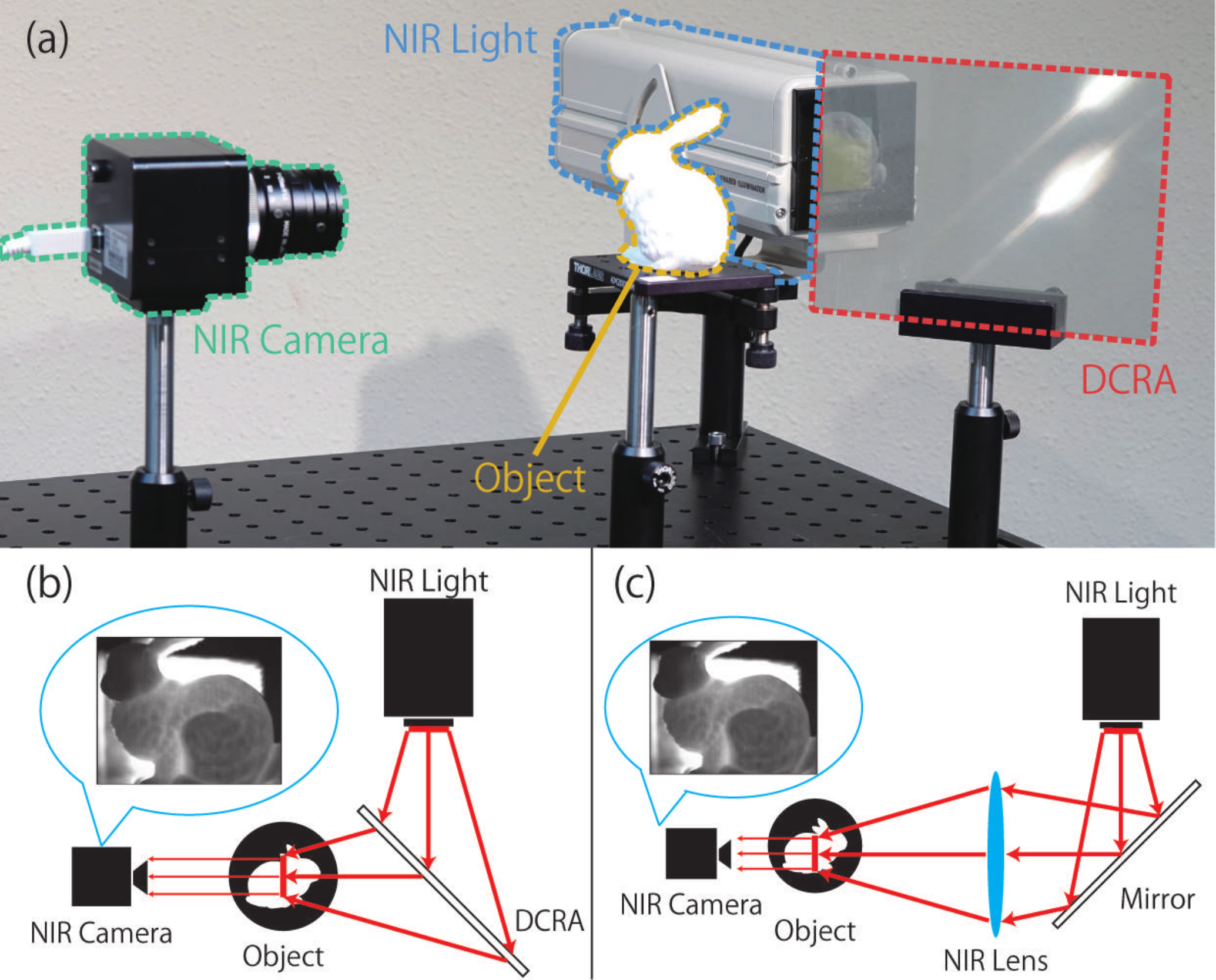}
    \caption{Optical System. (a) A picture of the imaging system used in this study. (b) A plan view of (a), showing the path of light. (c) Alternative imaging system.}
    \label{fig:fig03}
\end{figure}
An optical system using near infrared (NIR) light was constructed to obtain an optimal transmission image of the object. The target object, NIR camera, NIR light source, and an optical element were arranged as shown in Fig.\ref{fig:fig03}~(b). In order to eliminate the influence of speckle, the halogen lamp was used as the near infrared light source, which was equipped with an infrared transmission filter. A Dihedral Corner Reflector Array (DCRA)\cite{maekawa2006transmissive} was used to irradiate the back of the object.
As shown in Fig.\ref{fig:fig03}~(b), DCRA has the function of imaging the light source at a plane-symmetrical position.
Similar functions can be realized with mirror and lens as shown in Fig.\ref{fig:fig03}~(c).
In this experiment, because it was difficult to incorporate large mirror and lens, the amount of illuminating light was maximized by using DCRA.

\subsubsection{Transmitted Images}
In Section 3.2, it was explained that inner support structures differ in type, and that for a given support structure, the infill patterns, densities, and positions of structural elements are variable. In this section, the appearance of the transmission image is examined when the inner support structure is either changed by design, or left unchanged such that variation arises from manufacturing errors.

\begin{figure}[t]
    \centering
    \includegraphics[width=\linewidth]{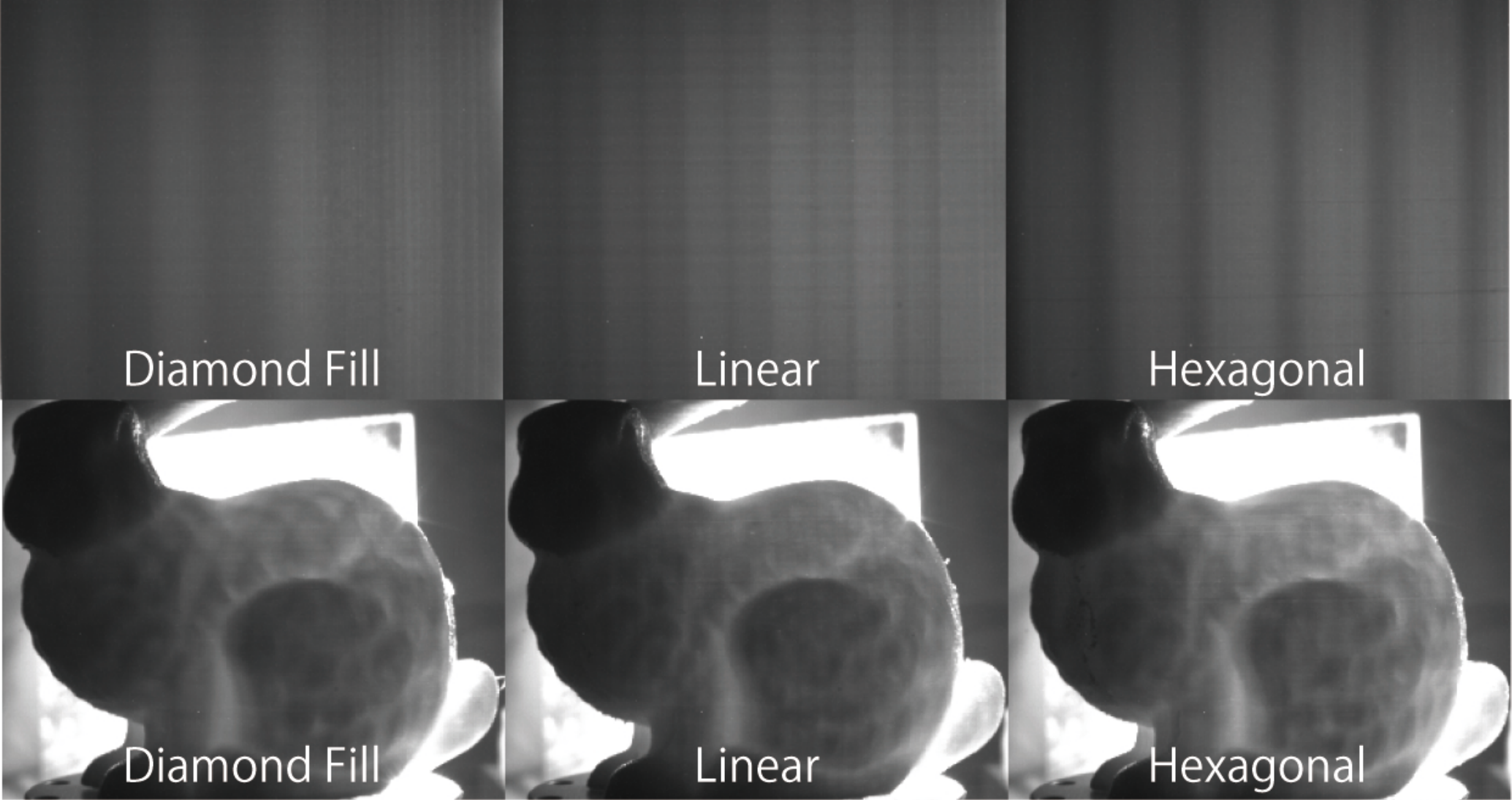}
    \caption{Variation of additive manufacturing infill pattern. (top) The transmission image when the infill pattern of the cube is changed. (bottom) The transmission image when the infill pattern of the bunny is changed.}
    \label{fig:fig04}
\end{figure}

{\bf Infill Pattern:}
Cube and bunny shapes were manufactured using three types of infill pattern: Diamond Fill, Hexagonal, and Linear. Transmission image results are shown in Fig.\ref{fig:fig04}. A different image was obtained for each infill pattern, for both cube and bunny. 

\begin{figure}[t]
    \centering
    \includegraphics[width=\linewidth]{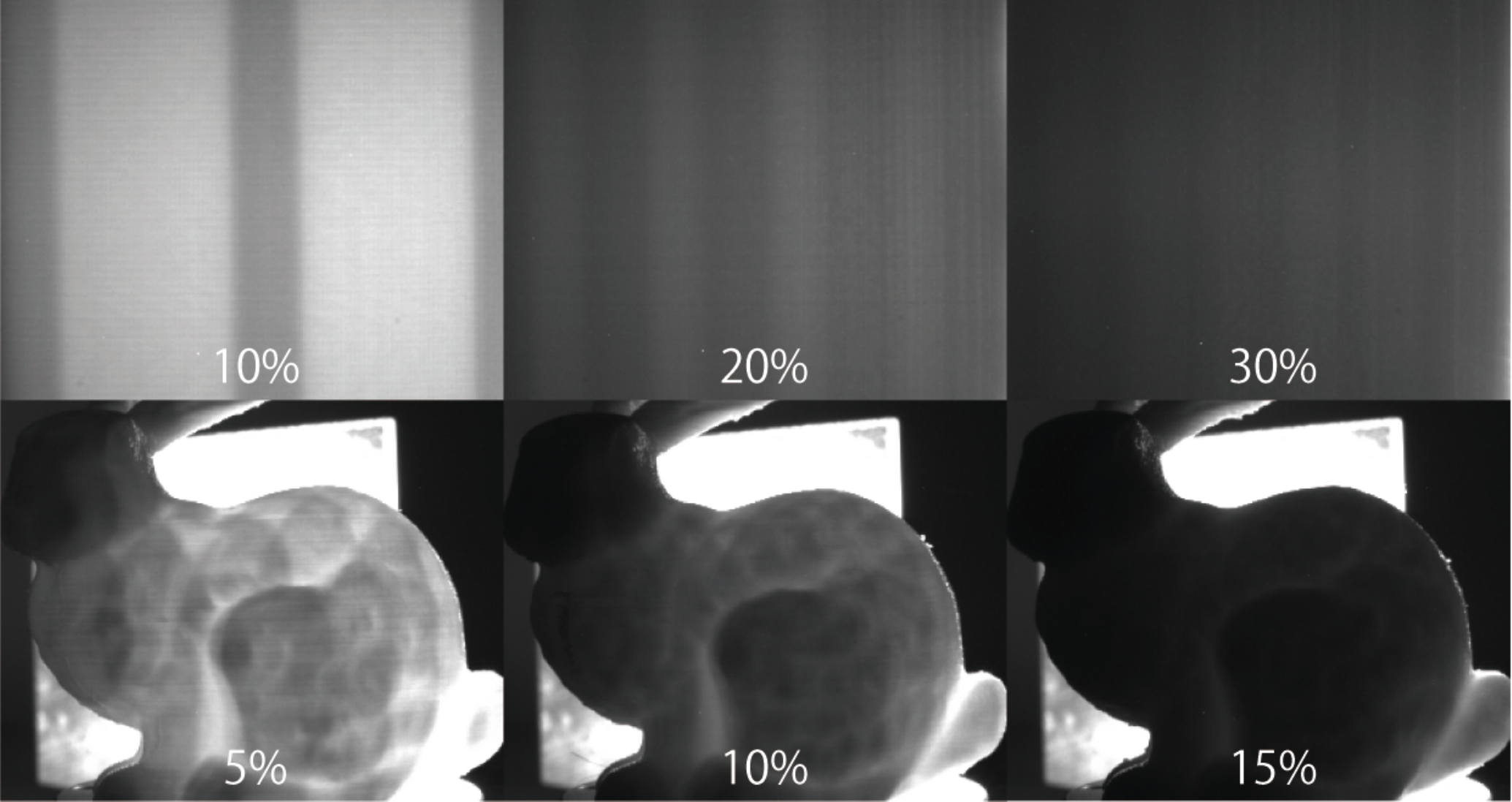}
    \caption{Variation of additive manufacturing infill density. (top) The transmission image when the infill density of the cube is changed. (bottom) The transmission image when the infill density of the bunny is changed.}
    \label{fig:fig05}
\end{figure}

{\bf Infill Density:}
Cube and bunny shapes were produced with Diamond Fill pattern densities of 5, 10, and 15\% for the cube, and 10, 20, and 30\% for the bunny. As the density increased, the amount of transmitted light decreased, hence the contrast in the transmissive images also decreased. These results are shown in Fig.\ref{fig:fig05}.

\begin{figure}[t]
    \centering
    \includegraphics[width=\linewidth]{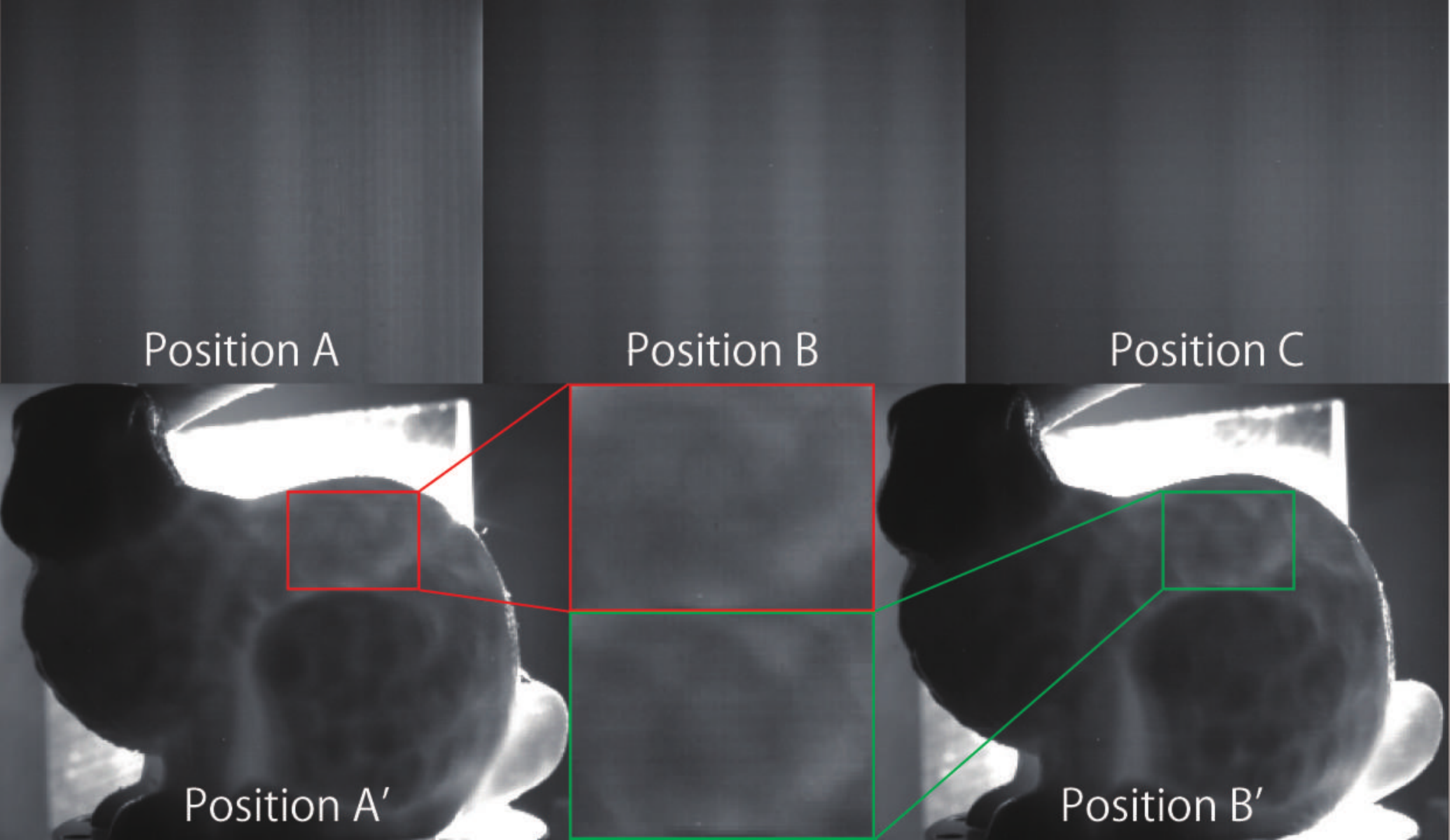}
    \caption{Variation of additive manufacturing infill position (top) The transmission image when the infill position of the cube is changed. (bottom) The transmission image when the infill position of the bunny is changed.}
    \label{fig:fig06}
\end{figure}

{\bf Infill Position:}
The same inner structure was created at multiple positions, with a Diamond Fill inner pattern density of 10\% for the cube, and 20\% for the bunny. Transmission image results are shown Fig.\ref{fig:fig06}. The cube images change as the position of the inner support structure changes. However, the transmission images of the bunny were almost unchanged due to the complexity of the model.

\begin{figure}[t]
    \centering
    \includegraphics[width=\linewidth]{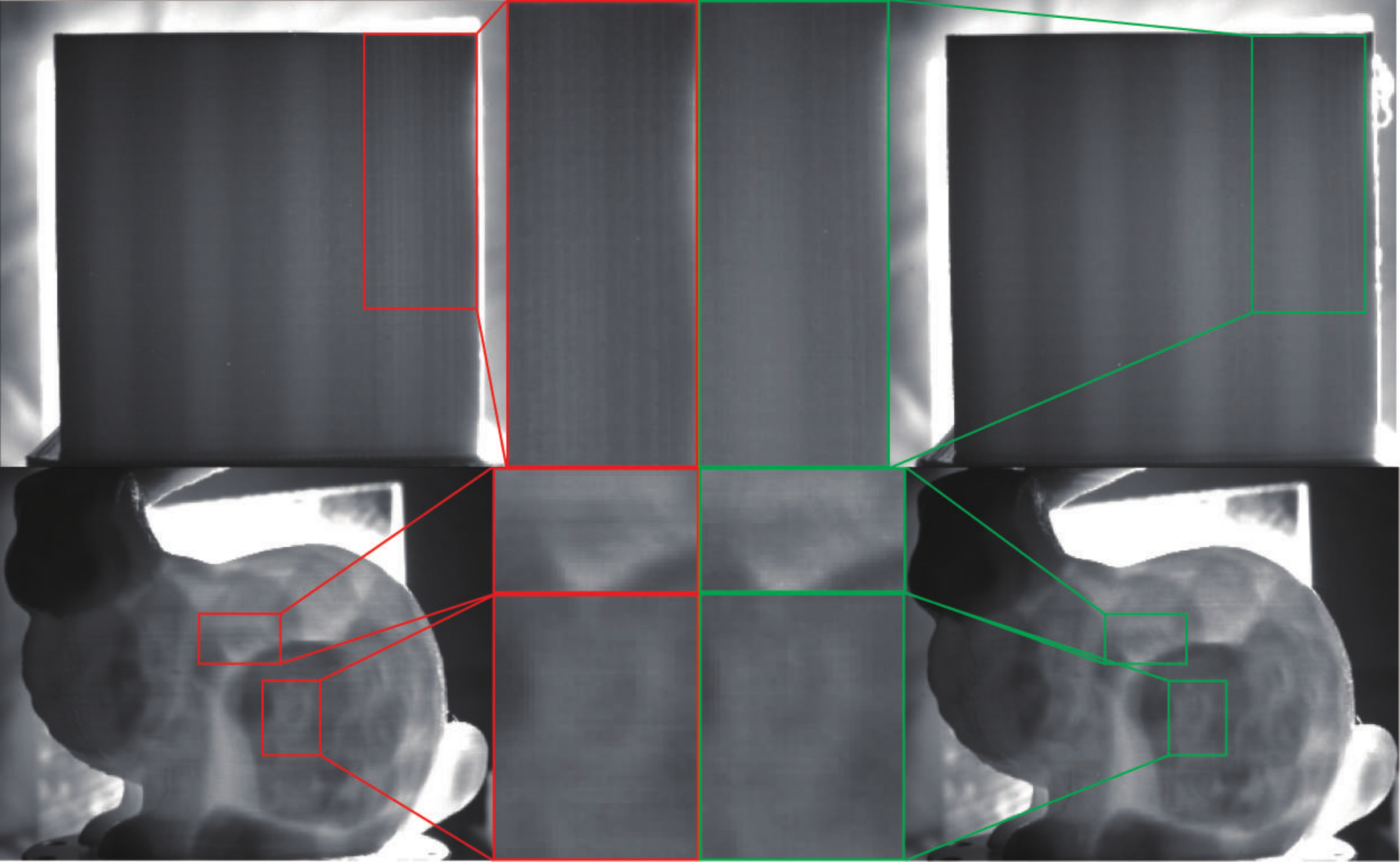}
    \caption{Random variation in items made with the same manufacturing parameters. (top) The comparison of transmission images of cubes with the same infill conditions. (bottom) The comparison of transmission images of additive manufactured bunnies with the same infill conditions.}
    \label{fig:fig07}
\end{figure}

{\bf Same Inner Parameters (only Manufacturing Errors):}
We compared the transmission images when all the parameters are equal.
In order to identify objects manufactured with the same parameters, it is necessary to utilize slight changes such as manufacturing errors.
We observed whether such a small difference could be acquired within a transmission image.
As shown in Fig.\ref{fig:fig07}, we prepared the cubes and bunnies with all same parameters listed in Table 2.
Since very similar images are acquired, the region of interest is enlarged.
In the enlarged image, it can be confirmed that trivial differences occur in both the cube and the bunny.

\section{Results}
\label{sec:experiments}
% \begin{figure*}[t]
\begin{figure}[t]
    \centering
    \includegraphics[width=\linewidth]{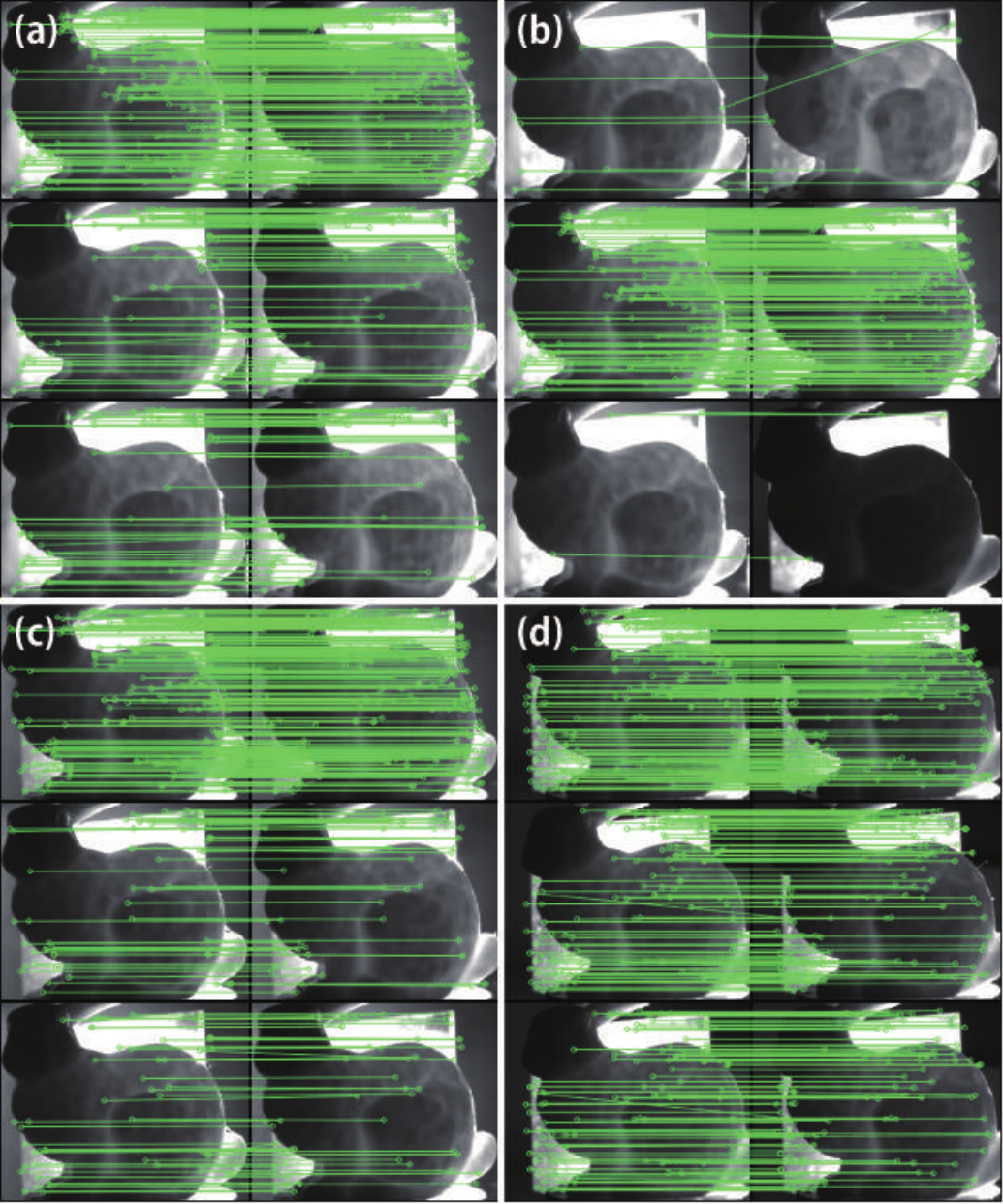}
    \caption{The results of feature matching when: (a) infill pattern is changed; (b) infill density is changed; (c) infill position is changed; and (d) comparing three bunnies with the same infill condition.}
    \label{fig:fig08}
\end{figure}
% \end{figure*}

In Section 3.3, the relationships between the characteristics of objects and their transmission images were established. When parameters related to shape or structure differ between objects, clear differences are likely to appear between their transmission images. However, only minor differences are visibly discernible between transmission images of objects with the same parameters. To evaluate the level of difference that can be resolved between various images, two algorithm-based experiments were conducted: feature matching, and classification by deep learning.

\subsection{Feature Matching}
To verify whether or not differences between the NIR transmission images were detectable by image processing, feature extraction and feature matching were initially conducted. In this experiment, the SURF algorithm\cite{SURF2006} was used to perform feature extraction.

The results of matching feature points are shown in Fig.\ref{fig:fig08}. Three key features were extracted for use in feature point matching. The matching features in a reference image and a target image were obtained by the k-nearest neighbor method. The D. Lowe ratio test\cite{Lowe2004} was used to select the top two matched features, the remaining feature is drawn in Fig.\ref{fig:fig08}. Feature matching was then performed between pairs of images that varied by infill pattern, or infill density, or infill position, or only by random differences due to manufacturing errors. Of the paired images showing matching results, the image on the left is the reference image, and the image on the right is the target image. The right hand images were changed and the results of feature matching were recorded as the number of matching points after the ratio test. These results are listed in Table 3. They quantify the change in matching with the change in a parameter; i.e., the degree to which the algorithm recognises the difference between images. A single reference image is used for all variations of a parameter hence the locations and maximum number of feature matching points are retained in the parameter series, thus allowing quantification and direct comparison of the results.

The results in Fig.\ref{fig:fig08} and Table 3 demonstrate that it is possible to discriminate between objects with obvious shape or structural differences by feature extraction and feature point matching using the SURF algorithm. This is not the case when objects are manufactured with the same structural parameters and differ only by manufacturing errors. Quantitative differences in matching results from one target image to the next were not significant, therefore feature matching is unlikely to be able to distinguish between objects at a higher degree of similarity.

\newcolumntype{M}[1]{>{\centering\arraybackslash}m{#1}}

\newcommand{\lightgray}{\cellcolor[rgb]{0.9, 0.9, 0.9}}
\newcommand{\middlegray}{\cellcolor[rgb]{0.75, 0.75, 0.75}}
\newcommand{\hardgray}{\cellcolor[rgb]{0.6, 0.6, 0.6}}

\renewcommand{\arraystretch}{1.15}
\begin{table}[t]
    \caption{Match rate of feature points. Since the same image is used to compare the same parameters, the match rate is 100\%.}
    \label{tab:feature_matching}
    \centering
    \footnotesize
    \begin{tabular}{|M{0.2\linewidth}|M{0.2\linewidth}|M{0.2\linewidth}|M{0.2\linewidth}|} \hline
        Infill Pattern \hardgray & Diamond \middlegray & Linear \middlegray & Hexagonal \middlegray \\ \hline
        Diamond \middlegray & 255 / 255 \lightgray & 82 / 255 & 62 / 266 \\ \hline
        Linear \middlegray & 79 / 261 & 261 / 261 \lightgray & 86 / 261 \\ \hline
        Hexagonal \middlegray & 67 / 305 & 81 / 305 & 305 / 305 \lightgray \\ \hline
        \hline
        Infill Density \hardgray & 10\% \middlegray & 20\% \middlegray & 30\% \middlegray \\ \hline
        10\% \middlegray & 310 / 310 \lightgray & 17 / 310 & 3 / 310 \\ \hline
        20\% \middlegray & 11 / 255 & 255 / 255 \lightgray & 3 / 255 \\ \hline
        30\% \middlegray & 2 / 174 & 5 / 174 & 174 / 174 \lightgray \\ \hline
        \hline
        Infill Position \hardgray & Position A \middlegray & Position B \middlegray & Position C \middlegray \\ \hline
        Position A \middlegray & 258 / 258 \cellcolor[rgb]{0.9, 0.9, 0.9} & 58 / 258 & 54 / 258 \\ \hline
        Position B \middlegray & 42 / 215 & 215 / 215 \lightgray & 35 / 215 \\ \hline
        Position C \middlegray & 49 / 265 & 39 / 265 & 265 / 265 \lightgray \\ \hline
        \hline
        Same Position \hardgray & Object 1 \middlegray & Object 2 \middlegray & Object 3 \middlegray \\ \hline
        Object 1 \middlegray & 235 / 235 \cellcolor[rgb]{0.9, 0.9, 0.9} & 128 / 235 & 107 / 235 \\ \hline
        Object 2 \middlegray & 122 / 227 & 227 / 227 \lightgray & 78 / 227 \\ \hline
        Object 3 \middlegray & 104 / 204 & 88 / 204 & 204 / 204 \lightgray \\ \hline
        % & & & \\ \hline
        % & & & \\ \hline
        % & & & \\ \hline
    \end{tabular}
\end{table}
\renewcommand{\arraystretch}{1.0}

\subsection{Deep Learning}
\begin{figure}[t]
    \centering
    \includegraphics[width=\linewidth]{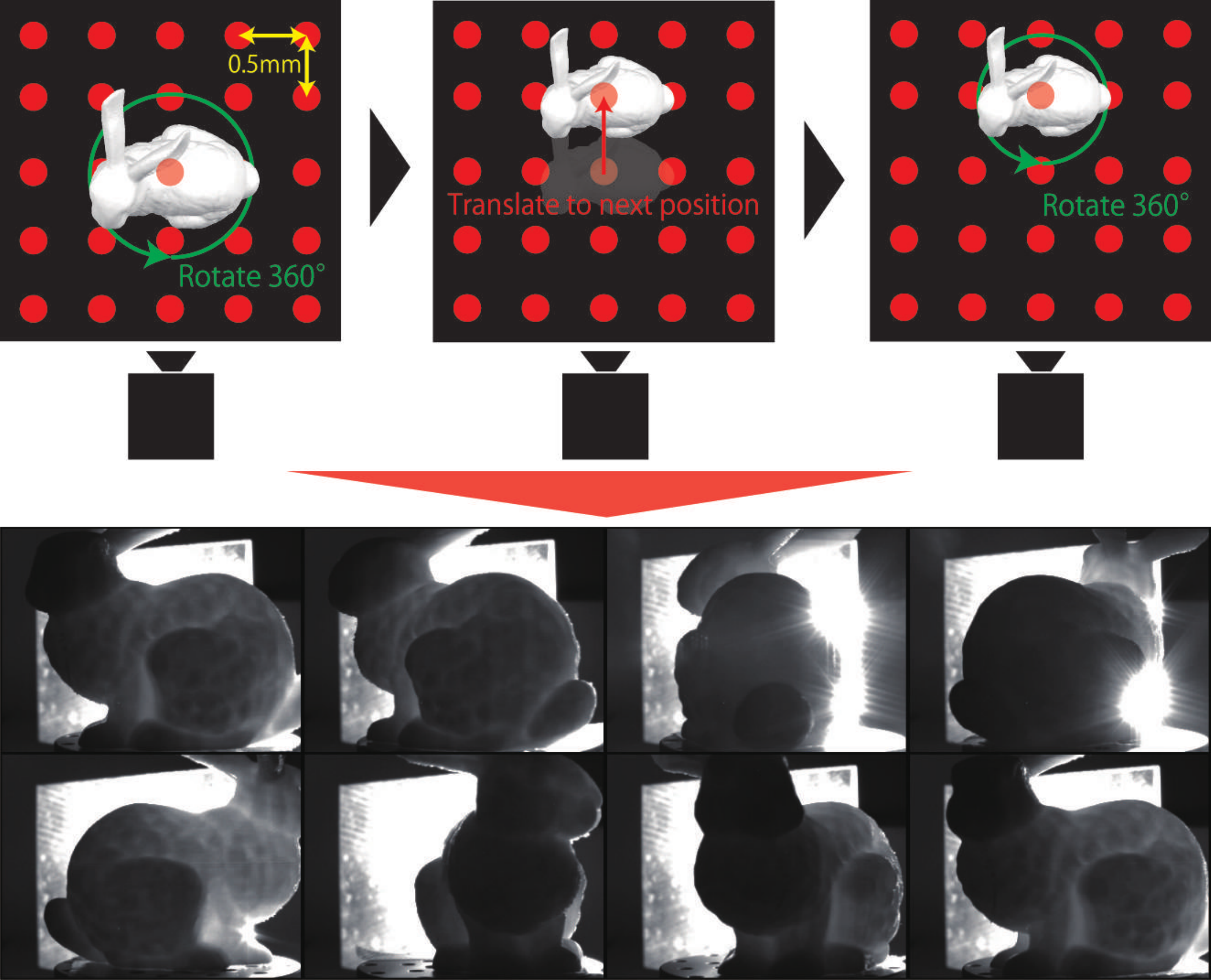}
    \caption{Flow of dataset creation for deep learning. (top) How to move an object when creating a dataset. (bottom) An example of a transmission image that was learned.}
    \label{fig:fig09}
\end{figure}

\renewcommand{\arraystretch}{1.2}
\begin{table}[t]
    \footnotesize
	\caption{Dataset Type.}
  	\begin{tabular}{|c|c|c|c|c|c|c|c|} \hline
     \lightgray & 3D Printer \lightgray & Position \lightgray & Pattern \lightgray & Density \lightgray & Amount \lightgray \\ \hline
    a & Makerbot & Random & Diamond Fill & 20\% & 10 \\ \hline
    b & Makerbot & Fixed & Diamond Fill & 20\% & 10 \\ \hline
    c & Makerbot & Fixed & Diamond Fill & 10\% & 10 \\ \hline
    d & Bellulo & Fixed & Honeycomb & 10\% & 10 \\ \hline
    % b & $\circ$ & Fixed & $\circ$ & $\circ$ \\ \hline
    % c & $\circ$ & $\circ$ & $\circ$ & 10\% \\ \hline
    % d & Bellulo & $\circ$ & Honeycomb & $\circ$ \\ \hline
  \end{tabular}
\end{table}
\renewcommand{\arraystretch}{1.0}

\begin{figure}[t]
    \centering
    \includegraphics[width=\linewidth]{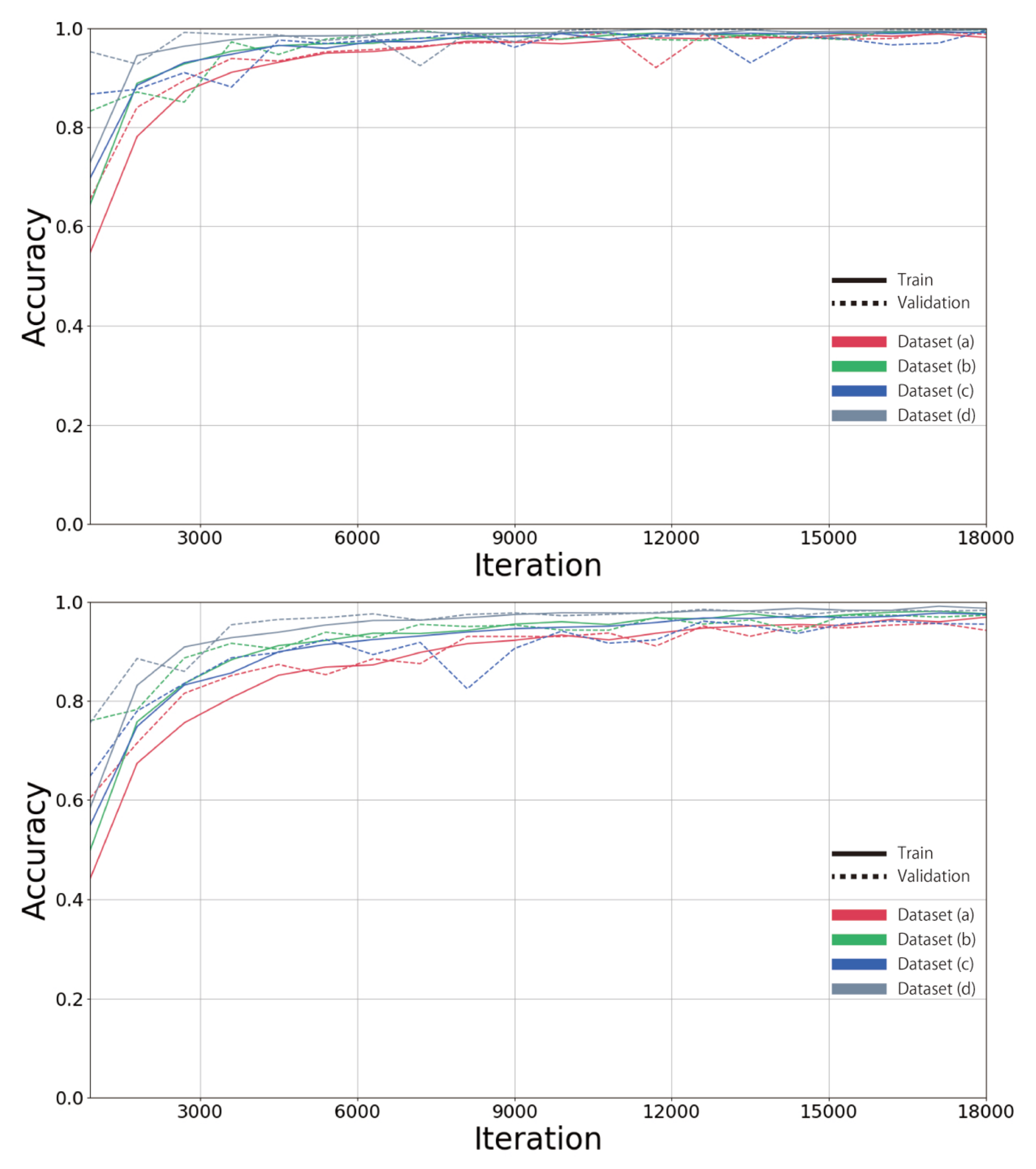}
    \caption{Deep learning second test. (top) Graph of learning progress when there is no noise in the dataset. (bottom) Graph of learning progress when noise is added to the dataset.}
    \label{fig:fig10}
\end{figure}

Deep learning has contributed greatly to the improvement of classification accuracy in recent years. Following the partial success of feature matching as discussed in the previous section, the deep learning technique was adopted to distinguish between items that are made using the same parameters, but differ due to manufacturing errors. Four groups of products were manufactured and their transmission images acquired. Datasets were produced from the transmission images (Table 4). Dataset (a) consists of products with different infill positions. Datasets (b), (c), and (d) consist of products that differ only as a result of manufacturing errors.

In order to apply the deep learning method, training datasets must first be prepared. The imaging system described in Section 3.3 (Fig. \ref{fig:fig03}) was used to obtain images for the datasets. In a commercial or industrial setting, the target may rotate slightly, or the set position can shift. To ensure that the identification system would be sufficiently robust to cope with this type of image variation, images with the target rotated and translated were included in the training dataset. As shown in Fig.\ref{fig:fig09}, 25 points were set at 0.5 mm intervals. By moving to each point and rotating 360 degrees at that location, images of objects transmitted at all angles were acquired. Transmission images were also obtained continuously, using the camera in video mode while the target was moving and rotating. Datasets were created by extracting an image from each frame of the video data.

Deep learning identification tests were performed using the datasets described above. Resnet\cite{He_2016_CVPR} was employed as the learning model and was trained in advance to improve learning efficiency.

Datasets for the first learning test consisted only of the data from standard transmission images; i.e., no added rotation or noise. The progress of the accuracy rate is shown in Fig.\ref{fig:fig10}(left). The accuracy of discrimination between similar images was high from the outset, due to the use of a trained neural network. As the classification rate reached 100\% in some cases, a more difficult classification task was set for the next test.

For the second learning test, the dataset included pre-processed images (augmented data), e.g., Gaussian noise and rotation images (5, 10, and 15 degrees). The progress of the accuracy rate is shown in Fig.\ref{fig:fig10}(right). Although the level of accuracy was lower than achieved in the first learning test, it was sufficient to distinguish between similar images. Hence, identification by deep learning is effective, even with noise added to the dataset.

\section{Discussions}
\label{sec:discussion}
The proposed system was able to identify products with high accuracy in the identification trials, however, there are still many points to be discussed. The limitations of the study are described below.  

\subsection{LIMITATIONS AND IMPROVEMENTS}

{\bf Transmitted Light and Contrast:}
The experimental investigation described in this paper was conducted on the premise that the transmitted light and contrast resolution, as available from the optical setup, were sufficient to acquire the characteristics of the objects being imaged. In future, it will be necessary to determine the minimum required transmitted light intensity, and the threshold of the contrast required to resolve all details in transmitted images.

{\bf Product Shape:}
As a preliminary investigation, this study was limited, on the basis of time and cost, to using a cube as a simple model and a bunny as a complex model. Further verification of the deep learning method will require resources for a large number of products to be manufactured with specified conditions, and their transmission images to be acquired as datasets. Products that do not have a flat bottom will also require a fixture to enable NIR photography.

{\bf Unverified Parameters:}
Changes to the “Inner Support Structure” sub-parameters, and manufacturing errors were the only product modifications considered in this paper. While differences arising from both were verified as recognizable by the identification system, the effects of changing the remaining unverified parameters must also be evaluated. “Printing Width” and “Layer Thickness” are additional shape-related parameters (Table 2). Neither was tested in this investigation as both can affect the overall appearance of the product. However, in some product designs it is possible to change these parameters without affecting the appearance of the product. The  effects of varying the parameters related to condition are also still to be tested.

{\bf Product Aging:}
In a real world distribution system, the product surface may be scratched or otherwise damaged during transportation, and product color can change over time. Although internal structure is less exposed to external impacts, the inner shape and the inner materials can undergo aging effects. The level of response of the detection system to such changes must be evaluated.

{\bf Optical System:}
The optical system is shown in Fig.\ref{fig:fig03}. Although DCRA was used to increase the amount of backlight in the small scale optical setup, the optical system should not require such special optical elements. In this study, the DCRA often appeared in transmission images (Fig.\ref{fig:fig08}). This problem needs to be addressed since transmission images are used as training data for the neural network, i.e. transmitted data must consist only of the object image, without any form of extraneous information. The optical system requires optimization in accordance with these issues.

{\bf Object Quantity:}
As shown in Table 4, 10 objects were prepared for each dataset.
This is because it took a considerable amount of time from the production of objects to shooting videos for datasets, hence it was difficult to prepare a large number of objects.
However, the actual item distribution system often handles more than 1000 items.
In the future, it is necessary to study how to deal with such a large amount of data.

{\bf Deep Learning Downside:}
Deep learning was adopted to realize high-precision identification. This method always creates the necessity for preparation of datasets and pre-learning. Dataset creation is also inefficient within the current system. The process requires simplification.

\begin{figure}[t]
    \centering
    \includegraphics[width=\linewidth]{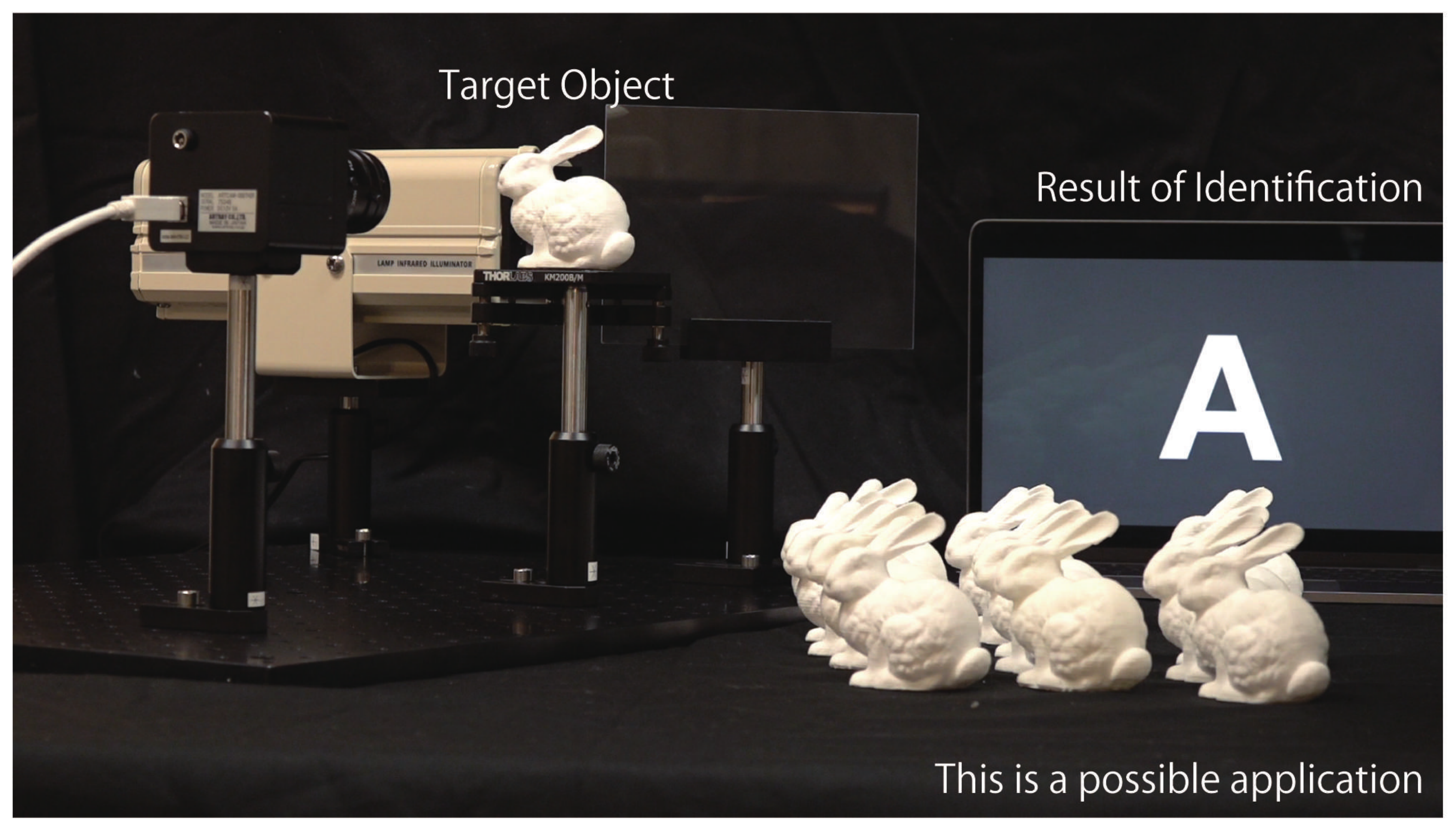}
    \caption{This is a future application that, like a point of sales (POS) system, identifies an object in real time and returns the result when the object is placed.}
    \label{fig:fig11}
\end{figure}

\section{Conclusion}
A product identification system using transmission images was developed. Product parameters relevant to the system were defined, and identification experiments were conducted based on a subset of several selected parameters. An optical system for acquiring transmission images was constructed, and the acquired transmission images were observed. Two methods, feature matching and deep learning, were verified as identification methods, and their functionality and limitations were analyzed.

Future experiments will be conducted under more complex conditions, using parameters not yet verified and objects of various shapes. Improvements will be made to the system, enabling it to operate effectively under many conditions. It will then be a useful system for identifying additively manufactured products to which the information embedding method cannot be applied (Fig.\ref{fig:fig11}).

\balance
\bibliographystyle{unsrt}
\bibliography{bibliography}

\appendix
\begin{figure*}[t]
    \centering
    \includegraphics[width=0.95\textwidth]{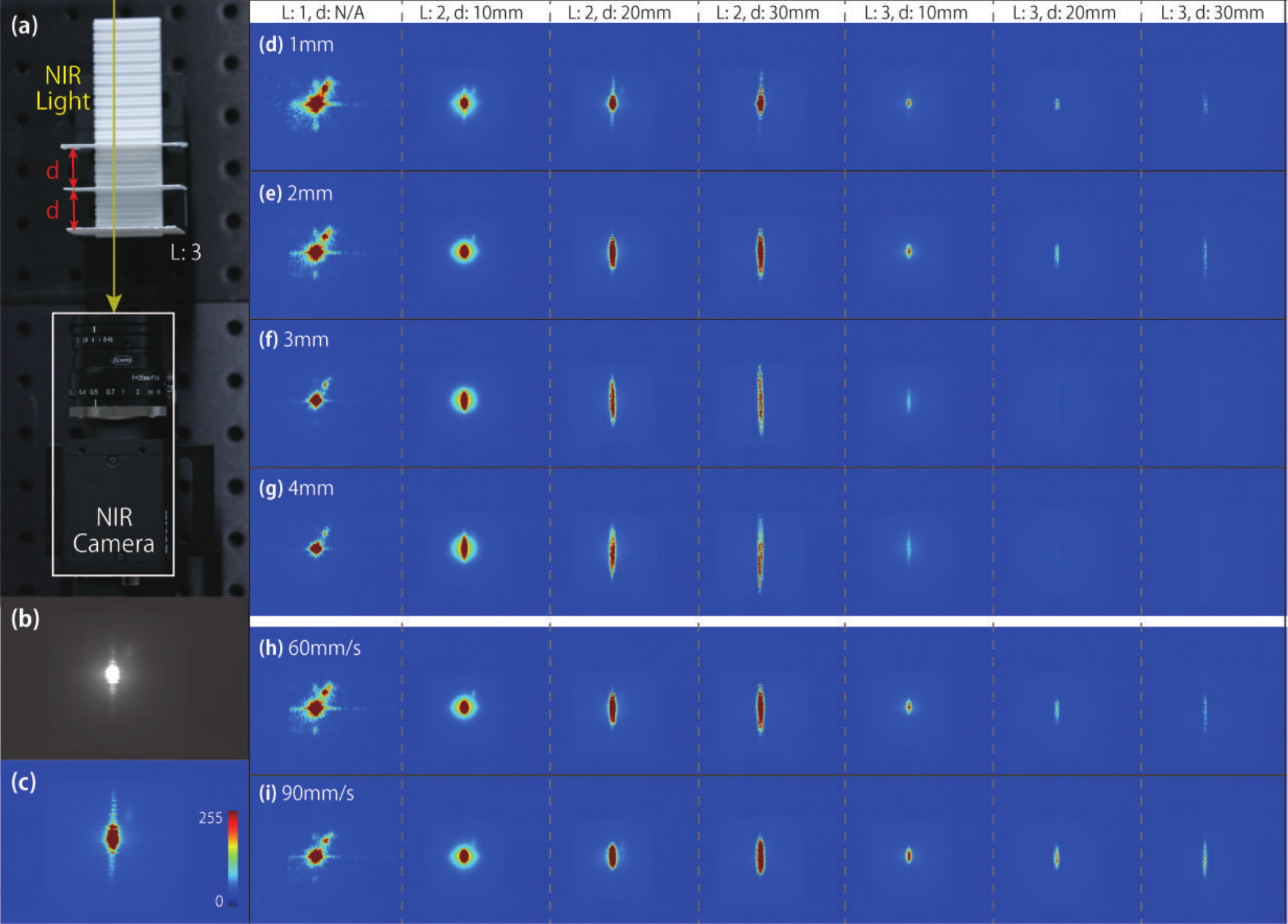}
    \caption{Changes in transmission image due to layer characteristics. (a) a photograph showing the experimental setup for imaging. (b) the actually acquired transmission image. (c) an image of (b) displayed in pseudo color. (d)-(g) the change of the transmission image when "Layer Thickness" is changed. (h), (i) the change of the transmission image when "Printing Speed" is changed.}
    \label{fig:one_layer}
\end{figure*}

\section{A Layer Feature}

\begin{figure*}[t]
    \centering
    \includegraphics[width=\textwidth]{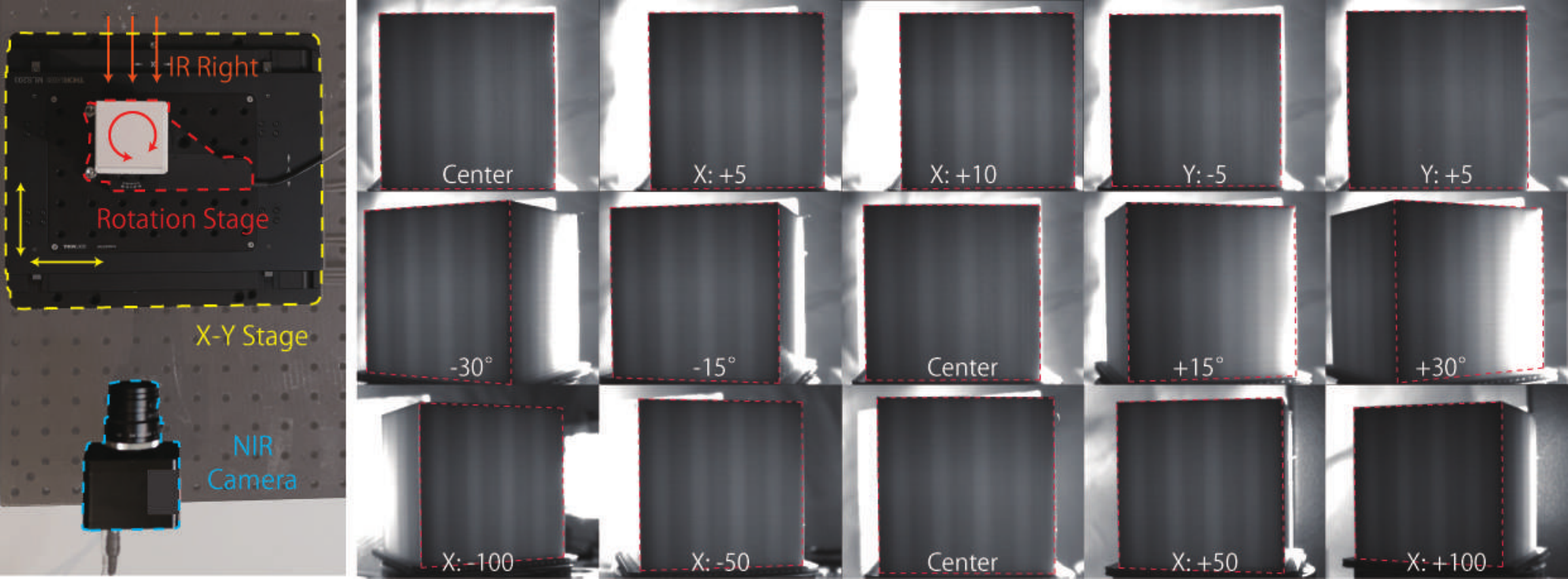}
    \caption{(left) a picture of an imaging system to verify robustness. (right top) the imaging results when the XY stage is moved in X and Y directions. (right middle) the imaging results when rotating the rotation stage. (right bottom) the result when the camera is moved in X direction and imaged toward the center of the object.}
    \label{fig:robust}
\end{figure*}

The transmission image of the product is a photograph of the light intensity emitted from a certain point after passing through various paths.
The smallest unit of a product that constitutes such a phenomenon can be defined as a "layer".
Regardless of the shape, infill pattern, infill density, and infill position, the transmitted light has only passed through many layers.
In other words, the stacked feature of each layer becomes a transmission image.
We have already defined the characteristics of each layer in Section 3.2.
Therefore, here we verified how the transmission image changes according to the change in the characteristics by changing the parameters of each layer.

The elements that make up the layer are broadly divided into two types: those that affect the shape of the product and those that affect the state of the product.
The elements that affect the shape are "Layer Thickness" and "Printing Width".
The elements that affect the state are "Printing Speed", "Temperature of Heat-Bed", "Temperature of Printing Head", "Retraction Distance", and "Environment".
The reason why "Inner Support Structure" is not included in these elements is that "Inner Support Strucure" is established by the positional relationship of multiple layers.

\subsection{Experimental Setup}
In order to confirm the effect on the transmission image of one layer, we installed a near infrared light source, a thin plate equivalent to one layer, and a near infrared camera as shown in Fig. \ref{fig:one_layer}(a).
To make it easier to see how the light source changed before and after passing through the layer, the shape of the light source entering the thin plate was made circular and the light source was collimated.
However, since coherent light is used as a light source, the acquired image is accompanied by speckles.
Therefore, in order to give priority to the measurement of the rough change of the light source shape, the influence of speckle was ignored and the saturation was allowed.
Also, in order to observe how the light diffuses after passing through one layer, it is necessary to take a transmission image after passing through multiple layers.
In addition, we prepared a foundation with grooves formed at intervals of 5 mm, and obtained a transmission image by fitting multiple sheets into the foundation while changing the distance between the sheets.

\subsection{Effect of Parameters}
Of the two types of elements that make up the layer, changes in shape are thought to have a large effect on the transmitted image.
Therefore, we first conducted an experiment to observe changes in the transmission image when the parameters about the shape were changed.
While "Printing Width" has the same value when using the same manufacturing equipment, "Layer Thickness" can be easily changed from the settings provided in the user interface. The transmission images were compared.
Fig.\ref{fig:one_layer}(b) shows what changes appear in the transmission image when the layer thickness is changed.
In this experiment, four layer thicknesses of 0.1 mm, 0.2 mm, 0.3 mm, and 0.4 mm were prepared.
We confirmed that the transmission image changed according to the thickness and positional relationship of the layer.

Next, an experiment was conducted to observe the change in the transmitted image accompanying the change in the parameters affecting the state.
"Printing Speed" was selected as an element related to the state.
The reason for verifying only one of the many state parameters is that the change in the state is considered to have a lower influence on the transmission image than the change in the shape.
Fig.\ref{fig:one_layer}(h) and (i) show the imaging results.
Compared to when the shape parameters were changed (Fig.\ref{fig:one_layer}(d)-(g)), the effect on the transmitted image was small.

As described above, we have compared the transmission images of elements related to shape and elements related to state.
Although there was a considerable change in both, the shape change is considered to be easy to use as a clue for identification.

\section{Robust Transmissive Images}
When using a transmission image for object identification, it is desirable that the transmission image is not affected even if the orientation of the backlight or the position and orientation of the object changes.
In order to achieve such robustness of the transmitted image, it is recommended that the transmitted light sufficiently passes through multipath.
In general, reflected light, internal diffusion, and transmitted light are generated when light enters an object.
If sufficient internal diffusion occurs, the transmission image obtained should be robust to the orientation and position of the target.

In order to verify the robustness of this system, we combined an XY scanning stage and a rotating stage to change the orientation and position of the target (Fig.\ref{fig:robust}(left)).
The XY scanning stage is Thorlabs MLS203-1, and the rotary stage is Thorlabs PRMTZ8.
In addition, for the object to be imaged, an object with a primitive shape was prepared: a cube with a size of 50x50x50 [mm], an infill density of 10\%, and an infill pattern of diamond fill.
Fig.\ref{fig:robust}(right) show the imaging results when the XY scanning stage is moved, the target is rotated, or the camera position is changed.
Looking at the region of interest in each imaging result, it was found that the orientation and position of the target had little effect on the transmitted image.

\end{document}